# Atomically Sharp Internal Interface in a Chiral Weyl Semimetal Nanowire

Nitish Mathur[1,2*], Fang Yuan[1], Guangming Cheng[3], Sahal Kaushik[4], Iñigo Robredo[5,6], Maia G. Vergniory[5,6], Jennifer Cano[7,8], Nan Yao[3], Song Jin[2*], Leslie M. Schoop[1*]

1. Department of Chemistry, Princeton University, Princeton, New Jersey 08544, USA
2. Department of Chemistry, University of Wisconsin-Madison, 1101 University Avenue, Madison, Wisconsin 53706, USA
3. Princeton Institute for Science and Technology of Materials, Princeton University, Princeton, New Jersey 08544, USA
4. Nordita, Stockholm University and KTH Royal Institute of Technology, Hannes Alfvéns väg12, SE-106 91 Stockholm, Sweden
5. Donostia International Physics Center, 20018 Donostia-San Sebastián, Spain
6. Max Planck Institute for Chemical Physics of Solids, Dresden D-01187, Germany
7. Department of Physics and Astronomy, Stony Brook University, Stony Brook, NY 11794, USA
8. Center for Computational Quantum Physics, Flatiron Institute, New York, NY 10010, USA

*email: nm8200@princeton.edu , jin@chem.wisc.edu, lschoop@princeton.edu

**ABSTRACT:** Internal interfaces in Weyl semimetals (WSMs) are predicted to host distinct topological features that are different from the commonly studied external interfaces (crystal-to-vacuum boundaries). However, the lack of atomically sharp and crystallographically oriented internal interfaces in WSMs makes it difficult to experimentally investigate hidden topological states buried inside the material. Here, we study a unique internal interface known as *merohedral* twin boundary in chemically synthesized single-crystal nanowires (NWs) of CoSi, a chiral WSM of space group $P2_13$ (No. 198). High resolution scanning transmission electron microscopy reveals that this internal interface is (001) twin plane and connects two enantiomeric counterparts at an atomically sharp interface with inversion twinning. *Ab-initio* calculations show localized internal Fermi arcs at the (001) twin boundary that can be clearly distinguished from both external Fermi arcs and bulk states. These merohedrally twinned CoSi NWs provide an ideal material system to probe unexplored topological properties associated with internal interfaces in WSMs.

Topology, a field of mathematics, has led to the prediction and realization of novel quantum materials which has reshaped the understanding of condensed matter physics and materials science.[1, 2] Topological quantum materials are distinguished from their topologically trivial counterparts by invariants derived from a material's bulk wave function. Weyl semimetals (WSMs), a subgroup of topological materials, have band structures made up of linearly dispersed electronic bands with topologically protected crossings (Weyl points) in three-dimensional momentum space, characterized by a topological invariant known as the Chern number ($C$).[3-5] The breaking of either time-reversal or inversion symmetry is crucial for the topological stability of a WSM state; it enables separation of the Weyl points of opposite chirality, leading to monopoles of Berry curvature in the momentum space.[3, 6] The non-trivial topology in WSMs is visible in their bulk-boundary correspondence, which manifests in the form of gapless topological surface states connecting the surface projections of bulk chiral Weyl points of opposite chirality, known as Fermi arcs.[3, 5, 6] These Fermi arcs can be experimentally observed *via* angle-resolved photoelectron spectroscopy (ARPES)[3, 7] and scanning tunneling microscopy (STM)[8-10] measurements on the clean material/vacuum boundaries (*external interfaces*) of WSMs. The bulk-boundary correspondence of WSMs can be further extended to *internal interfaces* such as twin/domain boundaries, which break the crystal lattice symmetry inside a material.[11] Despite several theoretical predictions on the possibility of localized Fermi arcs at the internal interfaces of chiral,[12] twisted,[13, 14] and magnetic WSMs,[15] there has not yet been experimental validation of the proposed internal Fermi arcs due to the lack of WSMs with well-defined internal interfaces.

In chiral WSMs, such as those adopting the cubic B20 structure ($P2_13$, space group 198), distinct internal Fermi arcs were proposed to exist at the twin boundary formed between crystals of opposite chirality (handedness).[12] The cubic B20 structure possesses a fixed chirality associated with the atomic arrangement. The space group symmetry results in multifold degenerate band crossings that feature large $C$ values.[16-23] Cubic B20 materials, including CoSi,[20, 21, 24, 25] RhSi,[20, 26] PtAl,[22] PtGa,[27] and PdGa,[23, 28] have been shown to host symmetry-enforced multifold band crossings at the high symmetry Brillouin Zone (BZ) points $\Gamma$ and $R$, which are maximally separated over a large energy window in the momentum space and connected by long Fermi arcs extending over the entire BZ. ARPES and STM measurements revealed that the sign of $C$ and the direction of Fermi-arc velocities associated with chiral fermions are reversed between the two structural enantiomers (left and right handedness).[22, 23, 28] The direct connection between the handedness (real space) and topological band structure (momentum space) of cubic B20 materials raises the question of the "bulk-defect plane" correspondence, i.e., how the bulk topology is reflected in the internal interfaces between the two enantiomers. This bulk-defect correspondence motivates experimental studies to detect hidden non-trivial topological states in the bulk of the material.

Although internal interfaces between the crystallites of opposite chirality can form during the growth of cubic B20 materials,[28-31] these interfaces are irregularly oriented, generally surrounded by other grain boundaries and do not extend throughout the bulk of the crystal. Possible WSMs with atomically sharp and crystallographically oriented internal interfaces would enable a deeper understanding of topological features and correlated phenomena at internal interfaces. Interestingly, single-crystalline nanowires (NWs) of FeSi and MnSi, which also have the cubic B20 structure, if synthesized *via* a "bottom-up" chemical vapor deposition (CVD) method,[32-35] often exhibit a unique internal interface that partitions the whole NW into two crystals of opposite chirality. Such interface, known as the merohedral twin boundary (Figure 1a), has been observed in NWs of FeSi[32] and MnSi[33] with cross-sectional transmission electron microscopy (TEM); however, both FeSi and MnSi have trivial band structures and do not host chiral fermions.[16] Cubic B20 CoSi is a well-known chiral topological semimetal with symmetry-protected multifold band crossings that have been experimentally observed.[20, 21, 24] In this letter, we present a CVD synthesis method to grow single-crystal NWs of cubic B20 CoSi that display merohedral twinning with an atomically sharp and crystallographically oriented internal interface. For the first time, we resolved the nature of merohedral twinning in cubic B20 CoSi NWs with high resolution scanning electron transmission microscopy (HRSTEM). Density functional theory (DFT) calculations further show the localized internal Fermi arcs at the (001) merohedral twin boundary of CoSi that are distinct from the external Fermi arcs and bulk states. This study provides an ideal material system to experimentally probe hidden topological states associated with internal interfaces in WSMs.

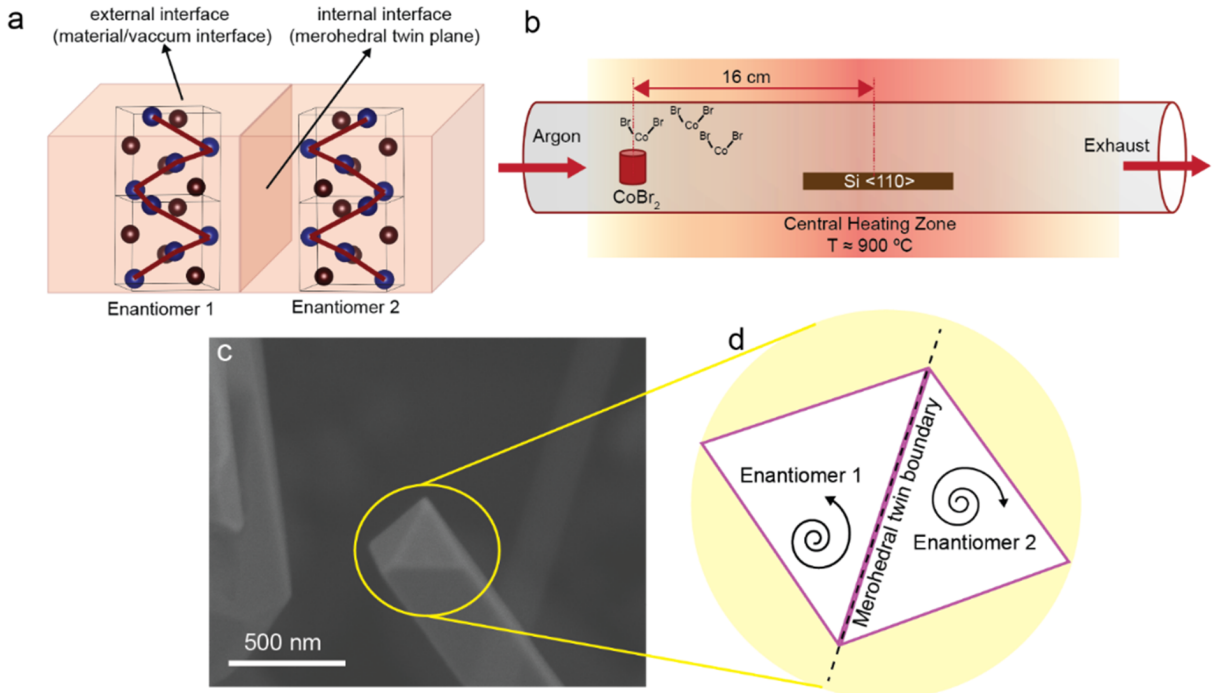

**Figure 1.** Growth and morphology of merohedrally twinned CoSi NWs *via* CVD. (a) A general scheme illustrating a merohedral twin plane connecting two structural enantiomers that resides in the bulk of the crystal, in contrast to the conventional external (materials/vacuum) interface. (b) Schematic of the CVD reaction setup for the synthesis of CoSi NWs. (c) High magnification SEM image of a merohedrally twinned CoSi NW displaying the characteristic rhombohedral tip morphology. (d) Schematic of the rhombus shaped cross-section of the representative faceted CoSi NW shown in panel c, depicting the formation of a merohedrally twinned boundary that partitions the whole NW into two structural enantiomers.

We performed the CVD reactions to grow CoSi NWs using pieces of Si (110) wafers covered with native $SiO_2$ layer as the growth substates and anhydrous $CoBr_2$ as the metal precursor in a custom-built CVD reactor (Figure 1b). The growth temperature of the Si substrate was maintained at 900 – 950 °C. After determining the temperature gradient of the single zone furnace along the length of the heated reaction tube, we placed the $CoBr_2$ precursor at 500 – 550°C, which was around 16 cm upstream to the central heating zone (see Supporting Information for details). CVD reactions conducted with the optimized parameters led to the formation of a variety of nanostructures or even micro-sized crystals on the substrates. We performed powder X-ray diffraction (PXRD) on post-growth and pre-growth (as reference) Si substrates to determine the phase of the CVD products (Figure S1). Apart from the peaks of Si substrate, all major diffraction peaks obtained from the area of high-density NW growth can be indexed to the cubic B20 CoSi phase with a lattice constant ($a$) of 0.445 nm. One impurity PXRD peak ($2\theta \sim 28.8°$) can be indexed to the cubic $\beta$-$CoSi_2$ phase (space group 225, $a$ = 0.536 nm) phase. Note that PXRD measurements can only provide average structural information about the products, which includes both nanostructures and bulk micro-sized crystals. Representative SEM images reveal different growth morphologies on the substrate, such as well-faceted thick NWs and nanorods (typical diameter ~ 200 nm – 1 μm) (Figure S2), thin NWs (typical diameter ~ 50 – 100 nm) (Figure S3a), and octahedral and cuboid shaped micro-crystals (> 1 μm) (Figure S3b). The impurity $\beta$-$CoSi_2$ phase revealed by PXRD comes from the cuboid shaped micro-crystals which show elemental composition of Co:Si ~ 1:2, while the octahedral shaped micro-crystals show Co:Si ~ 1:1 elemental composition (Figures S3c,d). Furthermore, high magnification SEM images of CoSi NWs clearly show well-faceted and smooth tips, which indicates that the NW growth does not go through the traditional vapor-liquid-solid (VLS) NW growth mechanism, since no particles were observed at the NW tips (Figure 1c).[36] Previous growth studies of cubic B20 silicide NWs suggests that the presence of a $SiO_2$ layer on the Si substrate could play a major role for the NW growth.[37-39]

Conveniently, it is simple to specifically pick the merohedrally twinned CoSi NWs amongst all the reaction products on the substrate by observing their morphology. A merohedrally twinned cubic B20 NW has a characteristic rhombohedral tip morphology, which is more clearly observed in SEM images of thicker CoSi (Figures 1c, more examples in Figure S2). The merohedral twin boundary resides inside the NW and divides it into two parts, which can be seen as two triangular facets at its tip (Figure 1d). Notably, this rhombohedral tip feature was also found in other merohedrally twinned cubic B20 NWs with a large diameter and the <110> growth direction, such as FeSi,[32] $Fe_{1-x}Co_xSi$[34] and MnSi.[33]

The high-resolution TEM (HRTEM) image of a representative merohedrally twinned CoSi NW also confirms the NW growth direction to be along <110> (Figure 2a). The corresponding fast Fourier transform (FFT) of the atomic-resolution HRTEM image shows peaks that can be indexed to the cubic B20 CoSi structure along the (1$\bar{1}$0) zone axis. The observed lattice spacings of 0.31 nm and 0.44 nm along (110) and (001), respectively agree with the theoretical values of the cubic B20 CoSi structure. Additionally, the entire surface of each NW is covered with a thin, amorphous $SiO_2$ layer (~ 3 nm) caused by surface oxidation. The quantitative SEM-EDS analysis on selected well-faceted merohedrally twinned CoSi NWs (Figure 2b, Figure S4 and Table S1) shows the elemental composition of CoSi NWs to be approximately 1:1 (Co:Si), which is within the instrumental error of standardless EDS analysis.[40] As there are no known polymorphs of CoSi monosilicide, we can conclude that these merohedrally twinned NWs adopt the cubic B20 structure.

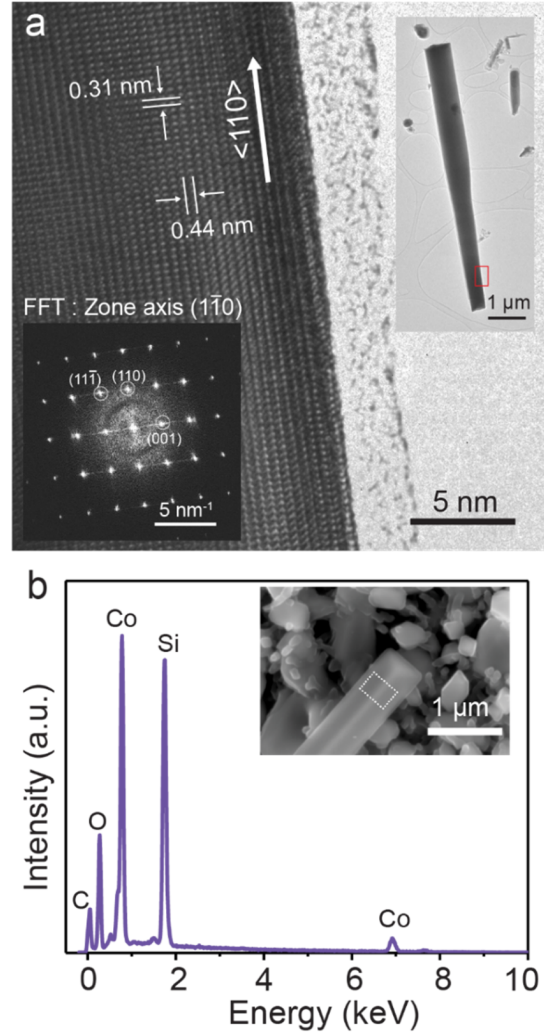

**Figure 2**. Characterization of merohedrally twinned CoSi NWs. (a) HRTEM image taken at the edge of a CoSi NW marked by the red box in the low magnification TEM image shown in the inset. The corresponding indexed FFT along the $(1\bar{1}0)$ zone axis determines the growth direction to be along <110>. The lattice spacings are marked along two planes in the perpendicular directions of (110) and (001). (b) Representative SEM-EDS spectrum recorded for a CoSi NW shown in the inset.

To investigate the nature and sharpness of the merohedrally twinned interface, we conducted cross-sectional HRSTEM. We picked a CoSi NW (thickness ~ 200 nm) with the merohedrally twinned facet lying flat on the Si substrate to facilitate the focused ion beam (FIB) cutting of a thin lamella. To avoid contamination or damage from the $Ga^+$ ion beam, a thick layer of amorphous carbon (1-2 μm) was deposited on the NW before FIB milling. A portion of the NW (~ 2 μm in length) was removed from the Si substrate and attached to a molybdenum TEM grid where the cross-section of the NW was further thinned down to < 100 nm using the FIB (see more details of the FIB process in Figure S5 in the Supporting Information). Finally, we collected TEM-EDS spectra on the cross-sectional TEM lamella to confirm its Co:Si ~ 1:1 elemental composition (Figure S6).

The twinning plane can be clearly observed in the TEM image as a contrast line running diagonally across the rhombus-shaped cross-section of the CoSi NW (Figure 3a). The corresponding selected area electron diffraction (SAED) pattern shows well-defined diffraction spots that can be indexed to the CoSi B20 phase along the (110) zone axis (Figure 3b), which agrees with the expected NW growth direction. A morphological analysis of the NW cross-section shows that the merohedral twin boundary (contrast line) and the crystal faces (rhombus side walls) belong to the {001} and {111} family of crystallographic planes, respectively. We further confirm the orientation of the twin plane and crystal faces by comparing the measured and calculated angles between the adjacent sides ($\theta_1$) and twin plane/sides ($\theta_2$) of the NW cross-section. The measured angles ($\theta_1 \sim 55°$, $\theta_2 \sim 110°$) match well with the calculated angles ($\theta_1 \sim 54.7°$, $\theta_2 \sim 109.5°$) (Figure 3a). We further employed high angle annular dark field (HAADF) STEM imaging to visualize the atomic arrangement at the merohedral twin boundary. The low magnification HAADF-STEM image clearly demonstrates the atomic arrangement of Co and Si atoms along (110) zone axis (Figure 3c). Additionally, an abrupt change in atomic arrangement (highlighted by the maroon stripe) near the twin boundary is clearly visible even in the low magnification HAADF-STEM image. The corresponding FFT (inset of Figure 3c) shows (001) twin plane in the NW.

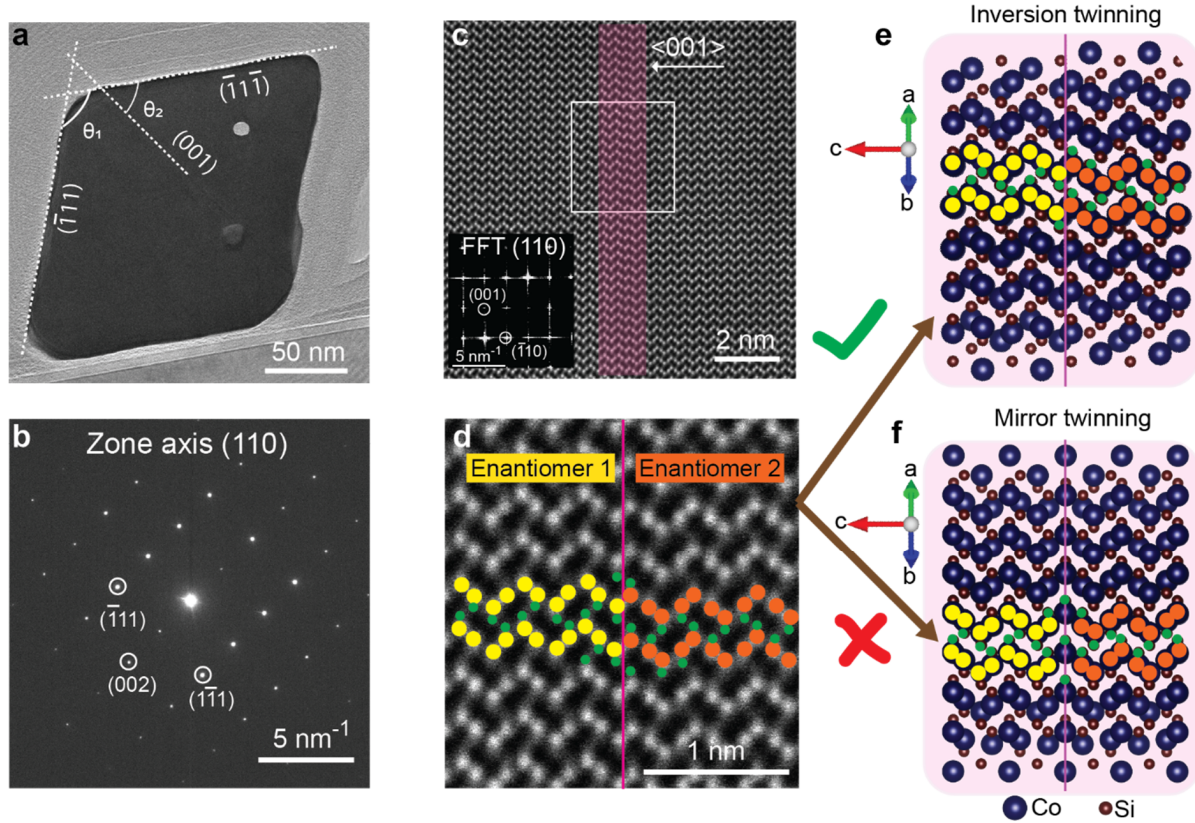

**Figure 3.** Cross-sectional HRSTEM imaging analysis of a merohedrally twinned CoSi NW. (a) TEM image displaying the FIB cut cross-section of a CoSi NW grown along the (110) crystallographic direction. (b) The corresponding SAED pattern is indexed to cubic B20 CoSi along the (110) zone axis. (c) Low magnification HAADF-STEM image; the maroon stripe highlights the abrupt change in atomic arrangement appearing due to the presence of a twin boundary. The inset of panel c shows FFT, which is indexed to cubic B20 CoSi along the (110) zone axis and determines the orientation of the twin boundary to be along the (001) plane. (d) High magnification HAADF-STEM image of the white rectangular box shown in panel c with the simplified structure models of two configurations of merohedrally twinned CoSi enantiomers superimposed. (e) Structural models of inversion twinning and (f) mirror twinning are compared with the high magnification HAADF-STEM image in (d). The maroon line in (d-f) marks the (001) merohedral twin boundary. In d-f, the yellow and orange circles highlight the Co atoms in two structural enantiomers i.e Enantiomer 1 and 2, respectively, and the green circles highlight the Si atoms.

A high magnification HAADF-STEM image (Figure 3d) of the area around the (001) twin boundary (enclosed in a white box in Figure 3c) suggests the chirality switching between two enantiomeric counterparts of the NW. The twin boundary is atomically sharp, as highlighted by the atomic arrangement of the Co and Si atoms in both enantiomers in Figure 3d. Merohedral twinning in a single crystal occurs when the twin operation (rotation, mirror, or inversion) belongs to a point group of the lattice vector but not of the crystal structure.[41] In merohedrally twinned crystals, the lattice of all twin components coincides exactly in both real and reciprocal space. Therefore, we observed no diffraction spot splitting in the SAED pattern (Figure 3b), as the enantiomeric counterparts of the merohedrally twinned CoSi NW completely overlap in reciprocal space. Here, we only considered twin operators, i.e., mirror and inversion that allow chirality flipping across the twin boundary of the NW. Due to the non-symmorphic space group of CoSi ($P2_13$), mirror and inversion twinning have different atomic configurations. The formation of a merohedral twinning plane in a cubic B20 crystal can be intuitively understood as gluing two crystals of opposite chirality in either of the two configurations where (1) the enantiomers share a plane that allows transformation into each other by mirror symmetry operation (mirror twinning) or (2) the enantiomers could be transformed into each other by an inversion center between two shared planes at the twin boundary (inversion twinning).[41, 42] Both inversion and mirror twinning are depicted in the simplified structure model of CoSi along (110) zone axis in Figures 3e and Figure 3f, respectively (see Supporting Information for detailed description of CoSi atomic models). The atomic arrangement displayed in the high magnification HAADF-STEM image (Figure 3d) can be perfectly traced atom-by-atom with the inversion twinning structure model (Figure 3e) as elucidated by highlighting atomic arrangement of Co and Si atoms. The coordination number of each Co and Si atom is preserved near the (001) twin boundary and no extra plane of atoms is observed at the twinning plane, which holds true only in case of the inversion twinning model (Figures S7,8). Additionally, for the inversion-twinned CoSi model, the external interfaces can be formed by the {111} family of planes as shown in Figure S9. We precisely observed this characteristic rhombus shaped cross-section in merohedrally twinned CoSi NWs (Figure 3a). This special case of merohedral twinning with an inversion twin operator is found only in non-centrosymmetric space groups and is known as racemic twinning.[42, 43]

Next, we qualitatively compare the merohedral twinned internal interface found in NWs of CoSi with other types of internal interfaces found during the growth of cubic B20 thin films and single crystals as potential material systems to detect topological features at the internal interfaces. In polycrystalline thin films of cubic B20 materials, the control over the selective growth of homochiral cubic B20 grains is difficult, and the final products generally have heterochiral growth domains with intergrowth regions of twins.[29, 30] Although internal interfaces such as twin boundaries are found between the adjacent grains with opposite crystal chirality in polycrystalline cubic B20 thin films, these internal interfaces are generally surrounded by other grain boundaries and do not extend throughout the bulk of the samples. In a cleaved (001) surface of cubic B20 single crystals, a structural domain wall boundary is occasionally formed between the crystal enantiomers at half step edges (step height ~ $a/2$), due to presence of a two-fold screw axis perpendicular to the surface.[28, 31] These domains can be transformed into each other by a glide-mirror symmetry operation. In contrast to the twinned internal interface, these domain boundaries in cubic B20 bulk crystals are irregularly oriented and do not extend throughout the bulk of the material, as these interfaces are only found at the step edges of a cleaved surface. The cross-sectional HRSTEM imaging of a single-crystal merohedrally twinned CoSi NW above clearly exhibits an atomically sharp internal interface that extends throughout the whole NW. The presence of clean and crystallographically oriented interfaces is crucial to study topologically protected surface states in WSMs.

Importantly, the (001) merohedral twin plane in CoSi NWs allows access to the unconventional multi-band crossings found in the electronic band structure (Figure 4a) at the $\Gamma$ and $R$ points of CoSi BZ (Figure 4b).[21, 24] We determined the Fermi-surface topology and stability of inversion-twinned (001) merohedral boundary of CoSi by performing DFT calculations as implemented in the Vienna ab initio simulation package (VASP) (see Supporting information for the details).[44-47] For these calculations, we considered a CoSi slab (Figure 4c, Figure S10a) periodic in $x$ ($a$-axis) and $y$ ($b$-axis) directions with a (001) twin boundary located at $z$ ($c$-axis) = 0.25 $a$ (Wyckoff position of CoSi shown in Table S2) and an inversion center located at (0, 0.4, 0.25) $a$. The external boundary of the CoSi 12.5(8.5)-unit cells slab is located at $z = \pm 6.25\ a$ ($\pm 4.25\ a$), starting from the twin boundary, respectively. Figure 4d and Figure S10b show Fermi arcs (internal and external) and bulk states from the Fermi surface calculations projected onto the area marked by the red box enclosing the (001) twin boundary at the Fermi energy for 12.5 and 8.5 CoSi unit cells, respectively. The intensity of the color scale in Figure 4d and Figure S10b represents the projection of internal (red) and external (blue) Fermi arcs. To elucidate the stability of the merohedral twin interface in CoSi, we determined the ground state energies of 12.5-unit cell models (Figure S11) with no internal interface ($E_{perfect}$) and (001) merohedral twin boundary ($E_{twin}$), respectively.

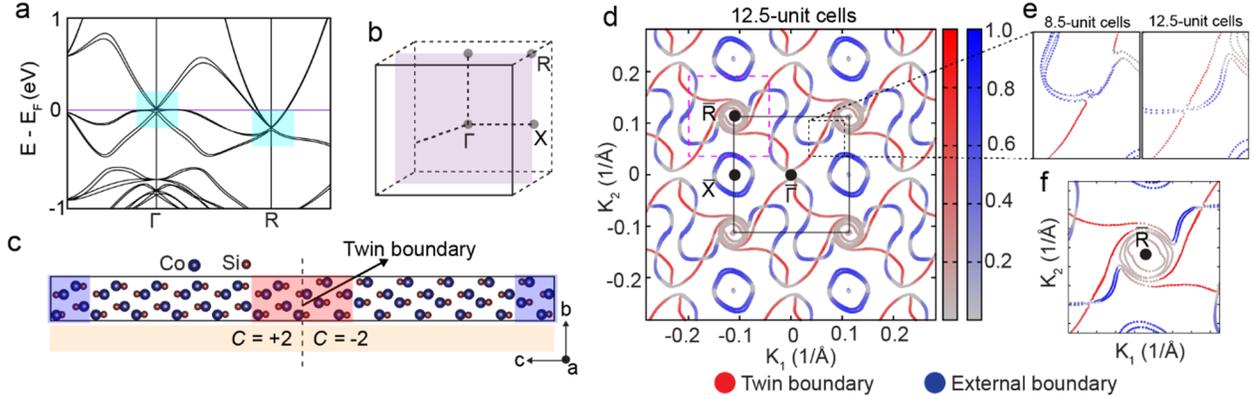

**Figure 4**. Fermi surface calculations for (001) merohedral twinned boundary of CoSi. (a) Electronic band structure of CoSi with multi band crossings at $\Gamma$ and $R$ points highlighted by cyan boxes. (b) Cubic BZ with (001) plane marked by a purple box. (c) A representative CoSi slab with 12.5-unit cells used in the DFT calculation, where the black dashed line represents the (001) merohedral twin boundary. Fermi surface calculations are projected onto the red box enclosing the twin boundary. The blue boxes at the end of the slab enclose external boundaries. The Chern number ($C$) of multi band crossings at $R$ is shown across the CoSi slab along <001> direction. (d) The calculated Fermi surface projection using the CoSi slab with 12.5-unit cells shown in panel c. The color scale (0-1) represents the projection of internal Fermi arcs (red), external Fermi arcs (blue) and bulk states (grey). The solid black box in panel d represents the BZ at the (001) twin boundary. (e) Zoom-in image of the area enclosed with dashed black box in panel d to illustrate hybridization between Fermi arcs in CoSi slab of 8.5 and 12.5-unit cells. (f) Zoom-in image of the area enclosing $R$ point marked by dashed maroon box. The internal and external Fermi arcs are highlighted in red and blue color, respectively.

We found that the difference in the ground state energies ($\Delta E = E_{\text{perfect}} - E_{\text{twin}}$) is notably low (0.196 J m$^{-2}$) when compared with the DFT calculated formation energies of twin boundary in elemental metals.[48] This could be caused by unchanged atomic coordination of Co and Si atoms at the merohedral twin boundary. These calculations suggest that the merohedral twin boundary is energetically stable once they form in the CoSi lattice. The internal Fermi arcs at the (001) twin boundary as shown in Figures 4d are easily distinguishable from the external Fermi arcs and bulk states, which means that the connectivity of internal Fermi arcs is different from the external Fermi arcs of both enantiomers. A recent theoretical study also shows that the connectivity of the internal Fermi arcs at a CoSi (001) twin boundary is different from the external Fermi arcs, and that it depends on the symmetry and energy of the internal interface.[12] Our calculation reveals that a significant dehybridization between the Fermi arcs occurs even with a slight increase in the unit cell size from 8.5 to 12.5, as shown in Figure 4e. This suggests that the extent of localization of the internal Fermi arcs can be enhanced by increasing the distance between the internal and external interfaces. The Fermi-surface topology shown in the case of 12.5-unit cells (Figure 4e) should better resemble the CoSi NWs synthesized herein (thickness > 50 nm), where we expect highly localized internal Fermi arcs at the twin boundary. Figure 4f shows four projected internal Fermi arcs (highlighted in red) and four external Fermi arcs (highlighted in blue) emanating from the R point, where the multifold band crossing has a $C = \pm 2$ (considering no spin-orbit coupling).[21] The change in $C$ ($\Delta C$) across the internal interface is equal to the number of gapless Fermi arcs on the surface BZ.[5] Figure 4c shows $C$ of multi band crossings at $R$ across the CoSi slab along the <001> direction. $C$ switches sign based on the handedness of the CoSi crystal.[22, 23] Hence, $\Delta C$ across the merohedral twin boundary is four, which results in a total of four internal Fermi arcs. Due to the small spin-orbit coupling of CoSi, we were unable to resolve the spin-split internal Fermi arcs in our calculations. In the case of cubic B20 materials that have substantial spin-orbit coupling to induce spin-split Fermi arcs, such as PdGa, a total of 8 distinct internal Fermi arcs can appear at the (001) merohedral twin boundary, as predicted in a previous study.[23] Transport measurements could unveil the topological connection of bulk Weyl bands and surface Fermi arcs in WSMs.[49, 50] Internal Fermi arcs in twinned Weyl semimetals are also predicted to exhibit transport signatures that should be visible in the wire's quantum oscillations or appear as a quantized chiral magnetic current.[12] It should be possible to fabricate nanodevices using the CoSi NWs reported herein to explore transport properties associated with internal Fermi arcs. Therefore, the topologically distinct phenomena exhibited by these highly localized internal Fermi arcs at the internal interface in comparison with the external interface of CoSi could be experimentally studied in the future.

In summary, we report the CVD synthesis of single-crystal NWs of B20 CoSi, a chiral WSM, with an atomically sharp merohedrally twinned internal interface. The characteristic rhombohedral tip faceting of these NWs enables convenient selection of merohedrally twinned CoSi NWs from the other reaction products. Cross-sectional HRSTEM imaging reveals an atomically sharp internal interface that lies at the intersection of two structural enantiomers of chiral B20 CoSi. Furthermore, careful analysis of atomic resolution HAADF-STEM images near the twin boundary reveals that it contains a center of inversion. The (001) merohedral twin plane in CoSi NWs allows access to the multifold band crossings at the $\Gamma$ and $R$ points in the BZ and their connecting long Fermi arcs. DFT calculations show distinguishable localized internal Fermi arcs at the (001) merohedral twinning plane. These arcs should be detectable with transport experiments performed on nanodevices based on these merohedrally twinned CoSi NWs. The availability of this nanomaterial system will facilitate experimental studies to detect and manipulate unexplored non-trivial topological phases hidden at the internal interfaces in topological materials.

## ASSOCIATED CONTENT

**Supporting Information**

The methods for CVD synthesis of CoSi NWs, material characterization and sample preparation, CoSi atomic models, and theoretical calculations; PXRD pattern, SEM images and SEM - EDS spectra of CoSi NWs and other reaction products; TEM sample preparation; CoSi atomic models used for DFT calculations and Fermi surface calculations.

**Notes**

The authors declare no competing financial interest

**ACKNOWLEDGMENTS**

This project is supported by the Gordon and Betty Moore Foundation's EPIQS initiative through award number GBMF9064, as well as the David and Lucille Packard foundation, the Alfred P. Sloan foundation, and the Princeton Catalysis Initiative (PCI). N.M., F.Y., G.C., N.Y., and L.M.S. acknowledge the sample characterization of the Imaging and Analysis Center (IAC) at Princeton University, partially supported by the Princeton Center for Complex Materials (PCCM) and the NSF-MRSEC program (MRSEC; DMR-2011750). N.M. and S.J. were supported by the NSF grant ECCS-1609585. M.G.V. and I.R. acknowledge the Spanish Ministerio de Ciencia e Innovación (grant PID2019-109905GB-C21), Programa Red Guipuzcoana de Ciencia, Tecnología e Innovación 2021 No. 2021-CIEN-000070-01 Gipuzkoa Next and the Deutsche Forschungsgemeinschaft (DFG, German Research Foundation) GA 3314/1-1-FOR 5249 (QUAST). J.C. acknowledges the support of the Flatiron Institute, a division of Simons Foundation, the Alfred P. Sloan Foundation through a Sloan Research Fellowship, and the US National Science Foundation under Grant No. DMR-1942447. S. K acknowledges the NordForsk which in part supports Nordita.